\documentclass[preprint,amsmath,amssymb,aps,showkeys,showpacs]{revtex4}

\usepackage[colorlinks=true, pdfstartview=FitV, linkcolor=blue, citecolor=red, urlcolor=magenta]{hyperref}
\usepackage[utf8]{inputenc}

\usepackage{multirow}

\usepackage{graphicx}
\usepackage{latexsym}
\usepackage{amsmath}
\usepackage{amsfonts}
\usepackage{amssymb}
\usepackage{verbatim}
\usepackage{dcolumn}
\usepackage{amssymb}
\usepackage{bm}
\usepackage{float}
\usepackage{subfigure}
\usepackage[english]{babel}
\newcommand{\be}{\begin{equation}}
\newcommand{\ee}{\end{equation}}
\newcommand{\bea}{\begin{eqnarray}}
\newcommand{\eea}{\end{eqnarray}}

\begin{document}

\newcommand{\NITK}{
\affiliation{Department of Physics, National Institute of Technology Karnataka, Surathkal  575 025, India}
}

\newcommand{\IIT}{\affiliation{
Department of Physics, Indian Institute of Technology, Ropar, Rupnagar, Punjab 140 001, India
}}

\title{Rotating 4D Gauss-Bonnet black hole as particle
accelerator}

\author{Naveena Kumara A.}
\email{naviphysics@gmail.com}
\NITK
\author{Ahmed Rizwan C.L.}
\email{ahmedrizwancl@gmail.com}
\NITK
\author{Kartheek Hegde}
\email{hegde.kartheek@gmail.com}
\NITK
\author{Md Sabir Ali}
\email{alimd.sabir3@gmail.com}
\IIT
\author{Ajith K.M.}
\email{ajithkm@gmail.com}
\NITK

\begin{abstract}
We demonstrate that the four-dimensional Gauss-Bonnet black hole can act as a particle accelerator with arbitrarily high centre-of-mass energy, during the collision of two general particles near the event horizon. The Gauss-Bonnet coupling constant $\alpha$, provides a deviation in the results from that of Kerr black hole. Our results show that the horizon structure, the range of allowed angular momentum and the critical angular momentum depend on the value of $\alpha$. For extremal cases, the centre-of-mass energy diverges near the horizon, suggesting that Gauss-Bonnet black hole can also act as a particle accelerator like a Kerr black hole. This is interesting in the context of probing the Planck scale physics. For the non-extremal case there exists a finite upper bound on the centre-of-mass energy, the maximal value of which depends on the parameter $\alpha$.

\end{abstract}


\maketitle


\section{Introduction}
It is well known that the extremal Kerr black hole can serve as a particle accelerator for the particles colliding in the vicinity of the black hole horizon \cite{Banados:2009pr}. The centre-of-mass energy of the colliding particles take arbitrarily high value at the horizon, provided one of the particle approaches with critical angular momentum. This fascinating physical process is called the BSW mechanism, which is a viable tool to probe the Planck-scale physics near the maximally spinning black holes. The mechanism is not only interesting from the theoretical perspective but also has implications on observational aspects like gamma ray burst and the active galactic nuclei. Therefore several studies soon followed to extend the BSW mechanism \cite{Berti:2009bk, Banados:2010kn, Jacobson:2009zg, Zaslavskii:2010aw, Wei:2010gq,Harada:2014vka}.  It has been shown that for a non-extremal Kerr black hole the centre-of-mass energy of the colliding particles is limited when the collision takes place at the inner horizon \cite{Lake:2010bq}. Later, the effect of the charge of the black hole on the centre-of-mass energy, in addition to the spin, was also investigated \cite{Wei:2010vca}. Wide variety of interesting applications of this mechanism exists in the literature \cite{Liu:2010ja, Mao:2010di, Zhu:2011ae, Zaslavskii:2012fh, Zaslavskii:2012qy, Zaslavskii:2010pw, Grib:2010xj, Harada:2011xz, Liu:2011wv, Patil:2010nt, Patil:2011aw, Patil:2011ya, Patil:2011uf, Amir:2015pja, Ghosh:2014mea} .

Recently there was a novel approach to probe the Gauss-Bonnet gravity in lower dimensions by D. Glavan and C. Lin \cite{Glavan:2019inb}. This theory is attractive as it has appealing features, like bypassing the Lovelock’s theorem and it is free from Ostrogradsky instability. Another important aspect of this theory is that, a positive Gauss-Bonnet coupling constant gives a static spherically symmetric solution, which is free from the singularity problem. The existence of such theories can be found even in the context of the conformal anomaly gravity  \citep{Cai:2009ua, Cai:2014jea}. The modified theory being in four dimensions, the interest stems from the astrophysical perspective as well. Several applications of the novel theory soon emerged in wide range of topics, like, the quasi-normal modes of scalar, electromagnetic and gravitational perturbations were studied in ref. \cite{Konoplya:2020bxa}. The stability and shadow of the black hole were also explored in the same article. Later, the innermost circular orbits and photon sphere of the black hole were studied using the geodesics \cite{Guo:2020zmf}. A wide variety of related topics were explored immediately after the proposal of this theory \cite{Casalino:2020kbt, Konoplya:2020qqh, Fernandes:2020rpa,  Lu:2020iav, Konoplya:2020ibi, Ghosh:2020syx, Konoplya:2020juj, Kobayashi:2020wqy, Zhang:2020qam, HosseiniMansoori:2020yfj, Kumar:2020uyz, Wei:2020poh, Churilova:2020aca, Islam:2020xmy, Liu:2020vkh, Konoplya:2020cbv, Jin:2020emq, Ai:2020peo, Heydari-Fard:2020sib, Li:2020tlo}. Recently, a rotating black hole solution was constructed for the novel Gauss-Bonnet Einstein gravity \cite{Wei:2020ght}. The focus was on the effect of the Gauss-Bonnet parameter on the shadow of the black hole. The rotating version of the black hole solution was also found in the ref. \cite{Kumar:2020owy}. In the present article we show that the rotating black hole in the Gauss-Bonnet-Einstein spacetime acts as a particle accelerator.

The article is organised as follows. In the following section (\ref{secone}) we discuss the properties of the black hole, mainly its horizon structure. In section \ref{sectwo} we focus on the orbit of a test particle in the vicinity of the black hole. In section \ref{secthree} the centre-of-mass energy of two general particles are examined, for both the extremal and non-extremal cases. We present our findings and discussions in section \ref{secfour}.


\section{Rotating 4D Gauss Bonnet black hole}
\label{secone}
\begin{figure}[H]
\centering
\subfigure[ref2][]{\includegraphics[scale=0.8]{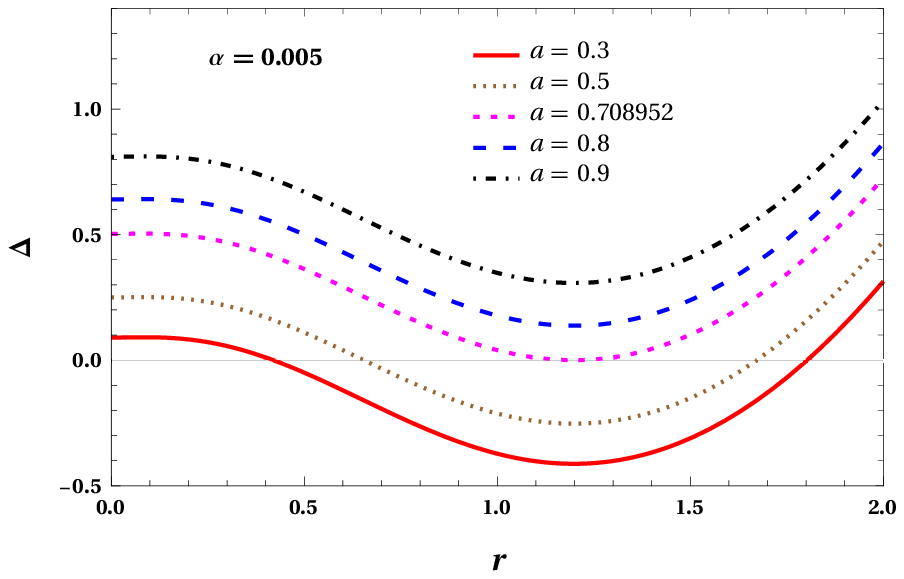}\label{horizon005}}
\qquad
\subfigure[ref1][]{\includegraphics[scale=0.8]{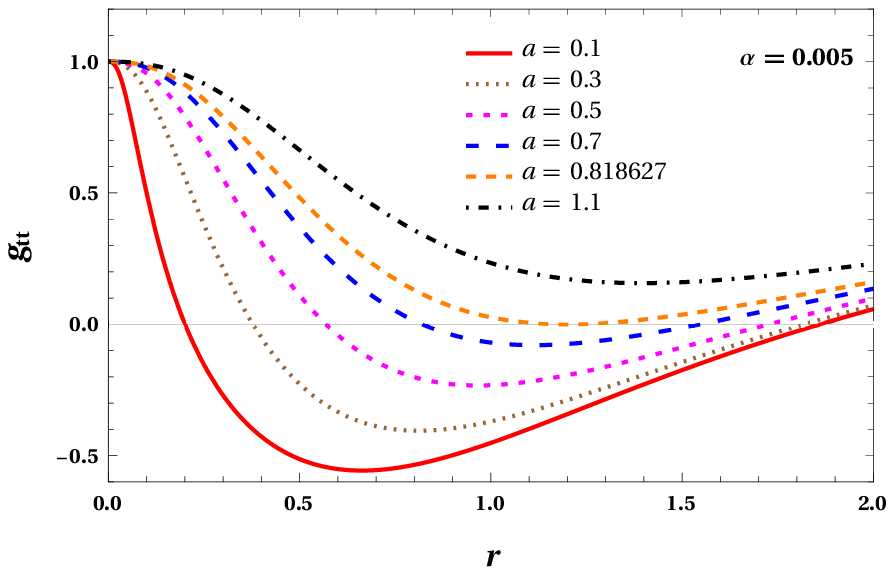}\label{sls005}}
\subfigure[ref2][]{\includegraphics[scale=0.8]{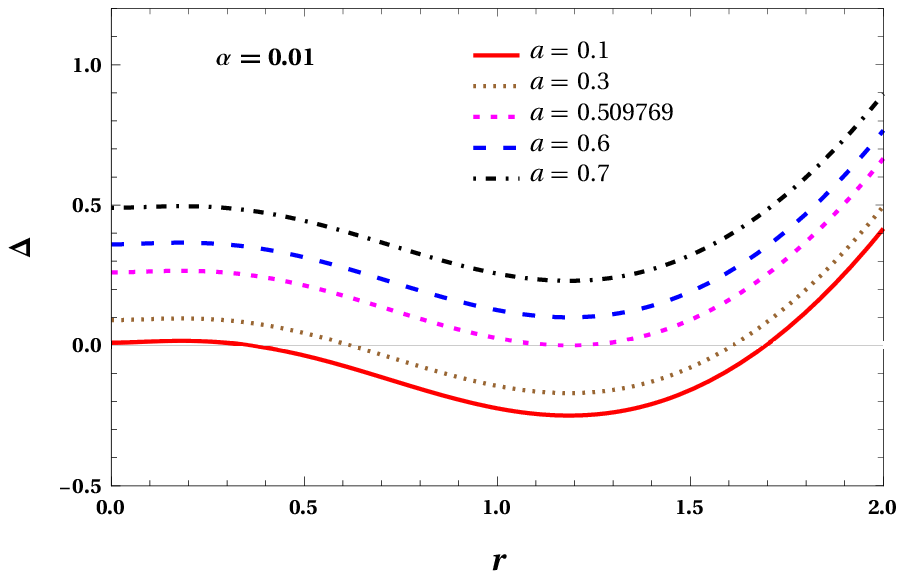}\label{horizon01}}
\qquad
\subfigure[ref1][]{\includegraphics[scale=0.8]{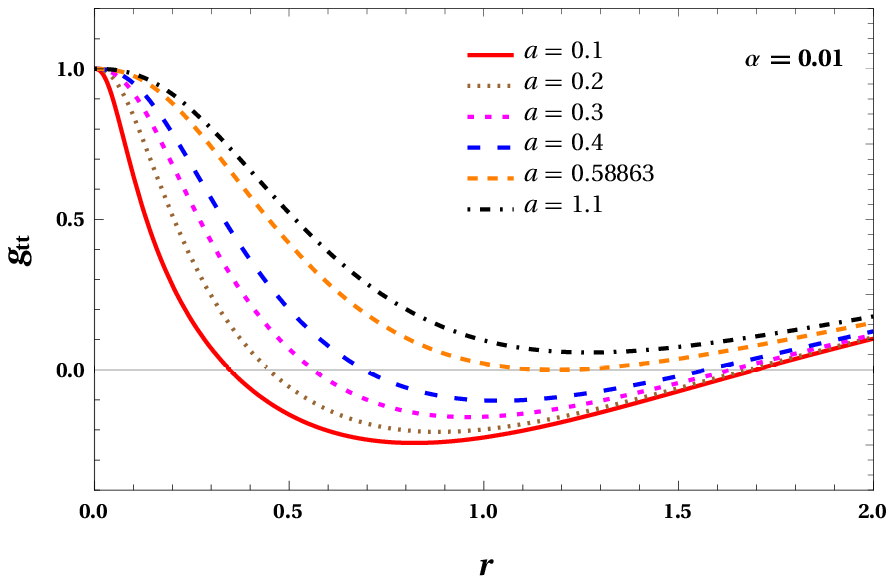}\label{sls01}}
\subfigure[ref2][]{\includegraphics[scale=0.8]{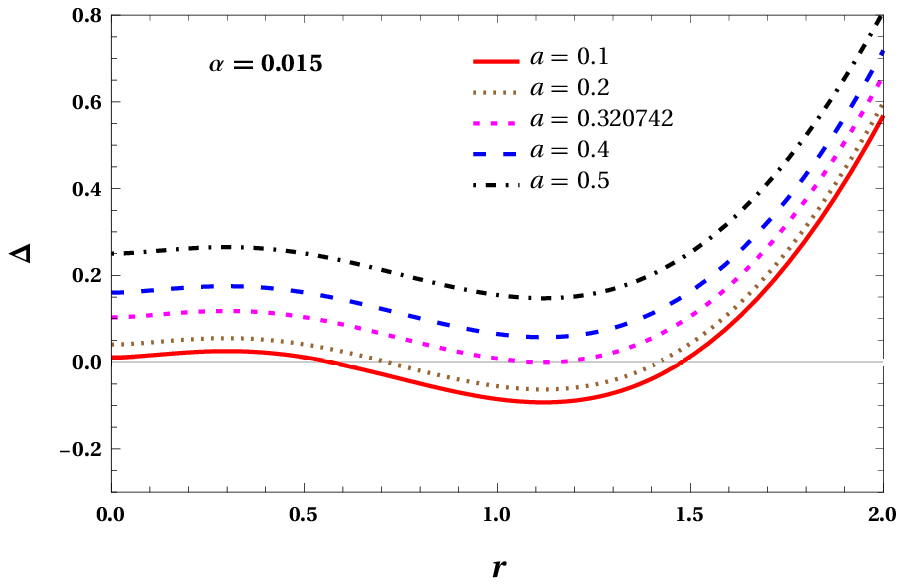}\label{horizon015}}
\qquad
\subfigure[ref1][]{\includegraphics[scale=0.8]{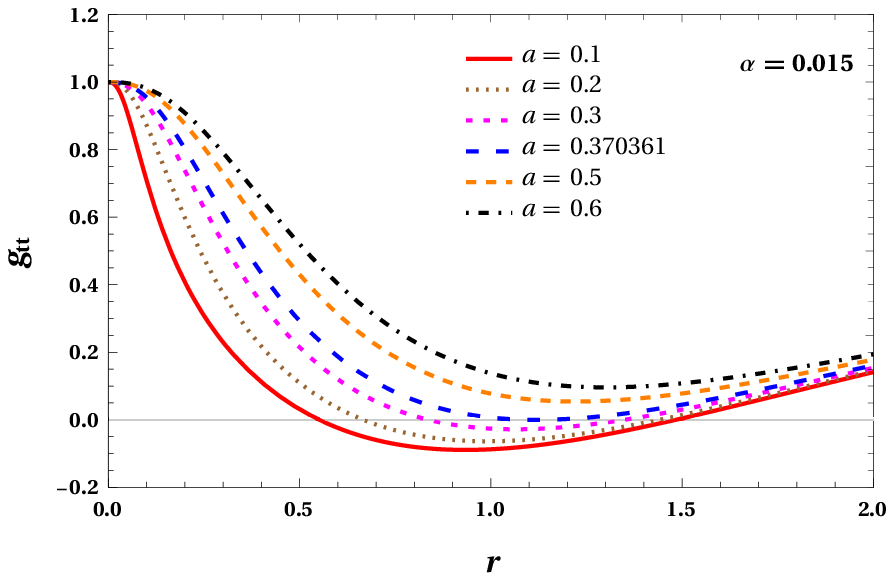}\label{sls015}}
\caption{Left panel : The horizon structure of the black hole for different values of $\alpha$. For a fixed allowed value of $\alpha$ there exist an extremal black hole with a single horizon (represented by the dotted (Magenta) line which just touches the x-axis). Right panel: The structure and location of the ergo surface of the black hole. The size of both the event horizon and ergo surface has a strong dependence on $\alpha$. The black hole mass is taken as $M=1$. In the ergo surface plots we have taken $\theta =\pi /6$.}
\label{horizon}
\end{figure}

In this article we aim to demonstrate that the four dimensional rotating black hole in the Gauss Bonnet gravity acts as a particle accelerator. In the Boyer-Lindquist coordinates, the stationary and axially symmetric (rotating) metric in this novel theory has the form \cite{Wei:2020ght, Kumar:2020owy},
\begin{eqnarray}
ds^2=&-\left( \frac{\Delta -a^2 \sin ^2\theta }{\Sigma } \right) dt^2 +\frac{\Sigma }{\Delta } dr^2-2-a \sin ^2\theta  \left(1-\frac{\Delta -a^2 \sin ^2\theta }{\Sigma }\right) dt d\phi\\ \nonumber
&+\Sigma d\theta ^2 +\sin ^2\theta  \left[a^2 \sin ^2\theta  \left(2-\frac{\Delta -a^2 \sin ^2\theta }{\Sigma }\right)+\Sigma \right] d\phi ^2,
\label{metric}
\end{eqnarray}
where

\begin{equation}
 \Delta =r^2+a^2+\frac{r^4}{32 \pi  \alpha } \left(1-\sqrt{1+\frac{128 \pi  \alpha  M}{r^3}}\right) \quad \text{ and} \quad   \Sigma =r^2+a^2 \cos ^2\theta .
\end{equation}
In the above expression, $M$ is the mass of the black hole, $a$ is the spin parameter and $\alpha$ is the Gauss-Bonnet coupling constant. Under the limit $a=0$ the solution (\ref{metric}) reduces to static spherically symmetric black hole whereas the limit $\alpha \rightarrow 0$ leads to the Kerr black hole. The rotating spacetime presented above is stationary and axisymmetric with Killing vectors
$\partial _t$ and $\partial _\phi$, corresponding to the time translation and rotational invariance, respectively. As in the case of Kerr metric, the rotating metric (\ref{metric}) is also
singular at $\Delta =0$. In general, the metric has two horizons, namely, the Cauchy horizon and the event horizon. The outer horizon, the surface of no return, is called the event horizon. We studied the horizon structure of the black hole for different allowed values of the coupling parameter $\alpha$ (table \ref{cauchyandevent} and the left panel of figure \ref{horizon}). The largest root of the equation $\Delta =0$ gives the location of the event horizon. The difference between the event horizon $r_H^+$ and the Cauchy horizon $r_H^-$ for varying spin parameter $a$. The size of the event horizon increases with $\alpha$ whereas the extent of the Cauchy horizon decreases, as a result the difference $\delta =r_H^+-r_H^-$ decreases accordingly. It is found that there exists an extremal value of $a=a_E$ for which the two horizons coincide, so that there are no roots at $a>a_E$. The black hole with $a=a_E$ is termed as an extremal black hole and with lower spin parameter than that of $a_E$ as a non-extremal black hole.

\begingroup
\setlength{\tabcolsep}{8pt} 
\renewcommand{\arraystretch}{1.2} 
\begin{center}
\begin{table}
\begin{tabular}{ |c|c|c|c|c|c|c|c| } 
\hline
\multicolumn{4}{|c|}{$\alpha =0.005$}&\multicolumn{3}{|c|}{$\alpha =0.01$}\\
\hline
$a$&$r_H^+$&$r_H^-$ &$\delta ^\alpha$ &$r_H^+$&$r_H^-$ &$\delta ^\alpha$\\
\hline
\hline
0.1 & 1.85861& 0.214307 &1.64431 &1.69559&0.36162&1.33397\\ 
0.2& 1.83826 & 0.316274 &1.52199 &1.66539&0.48145&1.18394\\ 
0.3& 1.80285 & 0.419869 &1.38298 &1.60977&0.61999&0.989783\\ 
0.4& 1.74947 & 0.53003 &1.21944 &1.51457&0.78801&0.726557 \\ 
$a_E$& 1.19825 & 1.19825 &0 &1.18569&1.18569&0\\ 
\hline
\end{tabular}
\caption{\label{cauchyandevent} The event horizon $r_H^+$ and the Cauchy horizon $r_H^-$ for the rotating 4D Gauss Bonnet black hole. The values given for different values of coupling constant $\alpha$ and their difference $\delta ^\alpha = r_H^+-r_H^-$ also depicted in the table.}
\end{table}
\end{center}
\endgroup

A spinning black hole also induces a frame-dragging effect on the fabric of space-time around it. This phenomenon leads to the creation of a cosmic whirlpool known as the ergosphere, which is located outside the event horizon. An object sitting inside the ergosphere is forced to move in the spinning direction of the black hole. An object falling into the ergosphere can escape the gravitational pull of the black hole with increased energy, where the additional energy is taken from the black hole. In this way, rotating black holes can act as powerhouses for the particles in its vicinity. To summarise, a rotating black hole has two horizon like surfaces: static limit surface and the event horizon. a rotating  The outer surface of the ergosphere is known as the static limit surface, the calculation of which requires the prefactor of $dt^2$ to vanish. In addition to the black hole parameters, the static limit surface also depends on the value of $\theta$ and coincide with the event horizon at the poles. For a fixed $\theta =\pi /6$ we have studied the location of ergo surface (static limit surface), which is depicted in the right panel of the figure (\ref{horizon}) for different values of $\alpha$ with varying spin parameter $a$. It is clear from the figure that the size of the ergosphere increase with $\alpha$. However, for a fixed value of black hole parameters $M$ and $a$, the static limit surface in this case is smaller than that of Kerr black hole. Nevertheless, the dependence of horizon structure and location of ergo surface on the Gauss-Bonnet coupling parameter shows that $\alpha$ has a significant effect on the energy extraction process.

\section{Orbit of the test particle around the black hole}
\label{sectwo}
In this section we the trajectory of a test particle of mass $\mu$ in the background of a rotating 4D Gauss-Bonnet black hole. We consider the particle motion in the equatorial plane described by the condition $\theta=\pi /2$. From the symmetry of the spacetime, the energy $E=-p_t$ and the angular momentum $L=p_\phi$ of the particle are conserved. These constants can be expressed as,
\begin{equation}
    E=-(g_{tt}\dot{t}+g_{t \phi }\dot{\phi})
\end{equation}
\begin{equation}
    L=g_{\phi \phi }\dot{\phi}+g_{t\phi}\dot{t}
\end{equation}
where the dot stands for the differentiation with respect to the affine parameter $\tau$ along the geodesics. The above equations can be reduced to the following forms,
\begin{equation}
\Sigma \frac{d t}{d \tau}=a(L-aE)+\frac{r^2+a^2}{\Delta}\left[ E(r^2+a^2)-aL\right]
\label{teqn}
\end{equation}
\begin{equation}
\Sigma \frac{d \phi}{d \tau}=(L-aE)+\frac{a}{\Delta}\left[ E(r^2+a^2)-aL\right].
\end{equation}
Now we focus on the circular orbits by analyzing the radial motion of the particle. The Hamilton Jacobi equation describing the particle motion is,
\begin{equation}
\frac{\partial S}{\partial \tau}=-\frac{1}{2}g^{\mu \nu} \frac{\partial S}{\partial x^{\mu}}\frac{\partial S}{\partial x^{\nu}}.
\label{jacobieqn}
\end{equation}
The Jacobi action $S$ can be separated as follows,
\begin{equation}
S=\frac{1}{2}\mu ^2\tau -E t +L\phi +S_r(r) 
\label{jacobiaction}
\end{equation}
where $S_r(r)$ is a function of the radial coordinate $r$. Substituting the action function \ref{jacobiaction} into the equation \ref{jacobieqn} and separating the coefficients of $r$ we get,
\begin{equation}
\Sigma \frac{d r}{d \tau}=\pm \sqrt{[(r^2+a^2)E-L a]^2-\Delta [\mu ^2r^2+\mathcal{K}+(L-aE)^2]}
\end{equation}
where $\mathcal{K}$ is the Carter constant. $\mathcal{K}$ is a constant of motion in addition to the energy $E$, angular momentum $L$ and mass $\mu$ of the particle. The Carter constant $\mathcal{K}$ is zero in for the particle motion along the equatorial plane. Not all the particles approaching the black hole will fall into the black hole. The fall or escape is determined by the impact parameter of the approaching particle. We examine the range of angular momentum, which is the window for the particle to fall into the black hole, by analysing the effective potential $V_{eff}$. The radial motion of the particle can be expressed as,
\begin{equation}
\frac{1}{2}\dot{r}^2+V_{eff}=0.
\end{equation}
where the effective potential is,
\begin{equation}
V_{eff}=\frac{\left[E \left(a^2+r^2\right)-a L\right]^2-\Delta \left[\mu ^2r^2+ \left(a E-L\right)^2\right]}{2 r ^4}.
\end{equation}
The circular orbits are characterised by the conditions,
\begin{equation}
V_{eff}=0\quad , \quad \frac{dV_{eff}}{d r}=0.
\end{equation}
These conditions give the range of angular momentum for the approaching particle. We have calculated the limiting values $L_{min}$ and  $L_{max}$ for different values coupling constant $\alpha$ for extremal (table \ref{lrangeext}) and non-extremal (table \ref{lrangenonext}) black holes. 
\begingroup
\setlength{\tabcolsep}{10pt} 
\renewcommand{\arraystretch}{1.2} 
\begin{center}
\begin{table}
\begin{tabular}{ |c|c|c|c|c|c| } 
\hline
$\alpha$ & $a_E$ & $r_H^E$ &$L_2$(min)&$L_1$(max)\\
\hline
\hline
0.005& 0.708953 & 1.19825 &-6.45388 &2.7342\\ 
0.0075& 0.605487 & 1.20044 &-6.23994 &2.97975\\ 
0.01& 0.509769 & 1.18569 &-6.03079 &3.25352\\ 
0.0125& 0.416597 & 1.15843 &-5.81545 &3.51396 \\ 
0.015& 0.320742 & 1.11961 &-5.58035 &3.78006\\ 
\hline
\end{tabular}
\caption{\label{lrangeext} The range of angular momentum of the infalling particle for the extremal rotating four dimensional Gauss Bonnet black hole.}
\end{table}
\end{center}
\endgroup

\begin{figure}[H]
\centering
\subfigure[ref2][]{\includegraphics[scale=0.8]{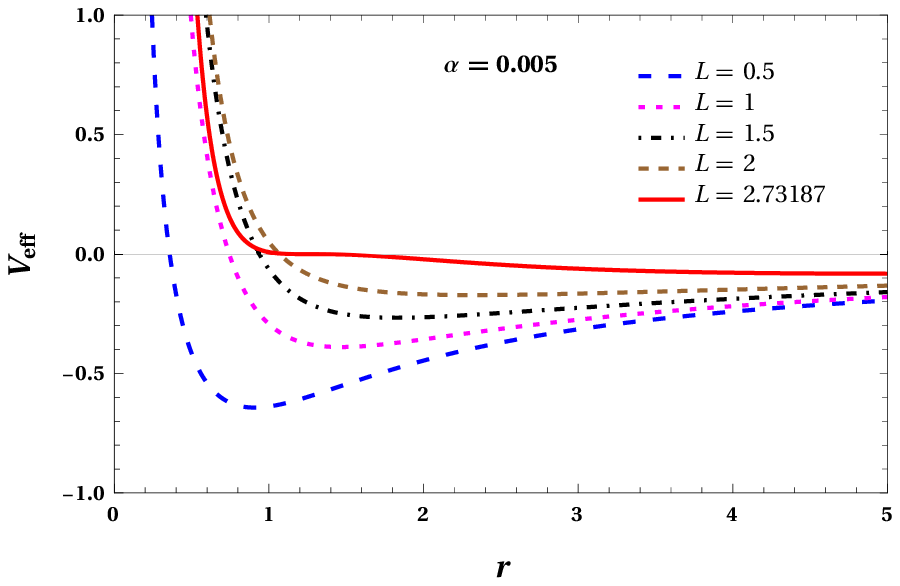}\label{Veff005a}}
\qquad
\subfigure[ref1][]{\includegraphics[scale=0.8]{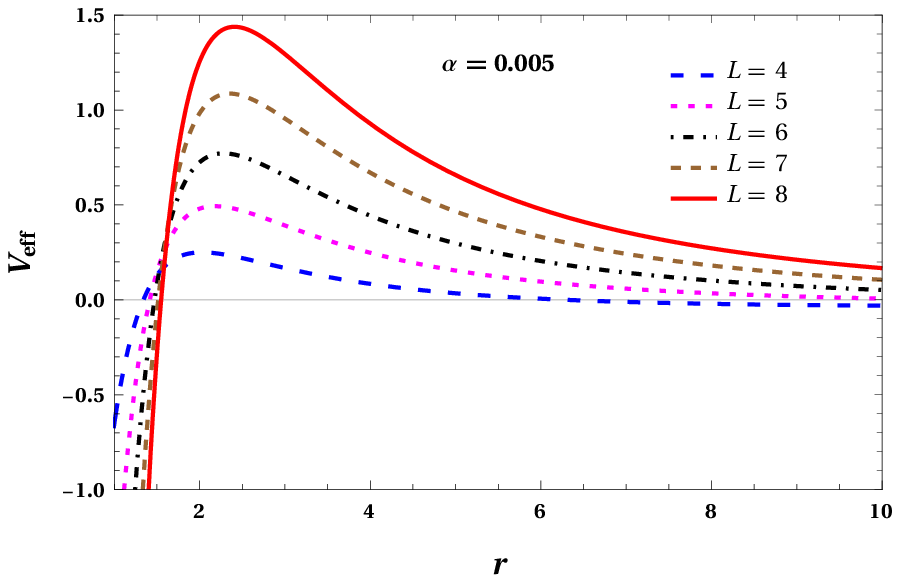}\label{Veff005b}}
\subfigure[ref2][]{\includegraphics[scale=0.8]{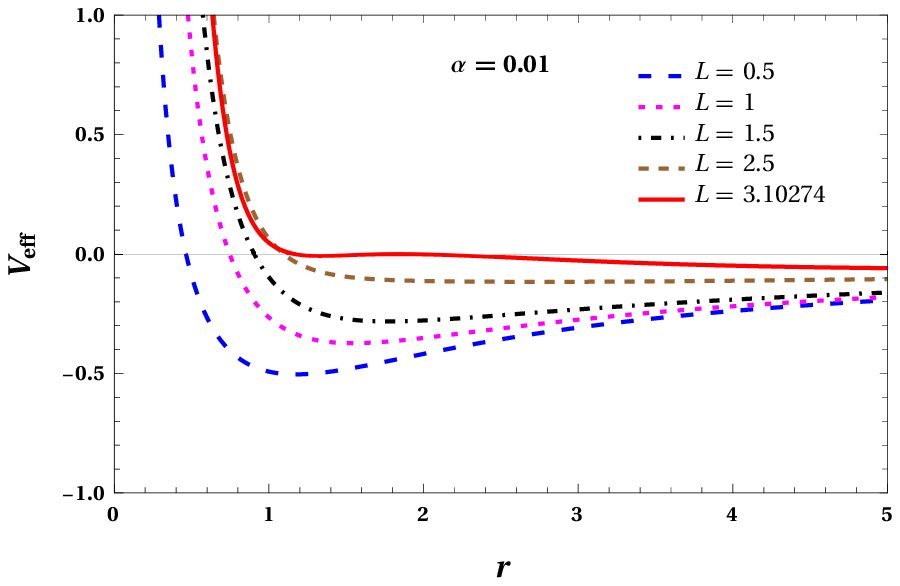}\label{Veff01a}}
\qquad
\subfigure[ref1][]{\includegraphics[scale=0.8]{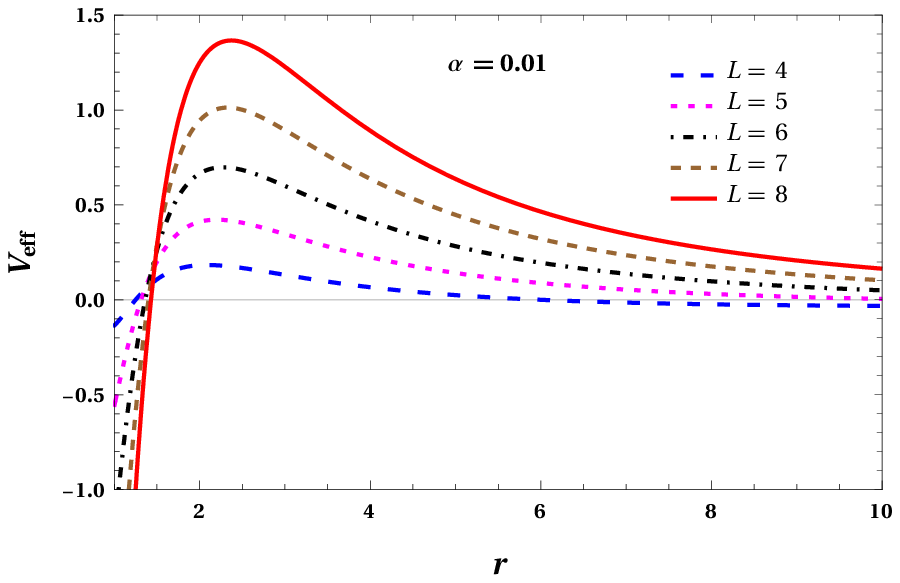}\label{Veff01b}}
\subfigure[ref2][]{\includegraphics[scale=0.8]{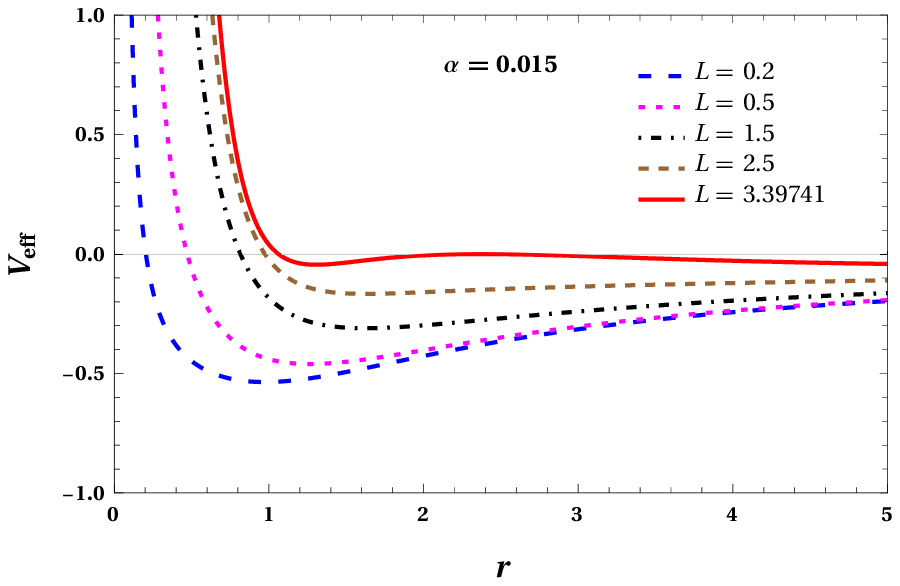}\label{Veff015a}}
\qquad
\subfigure[ref1][]{\includegraphics[scale=0.8]{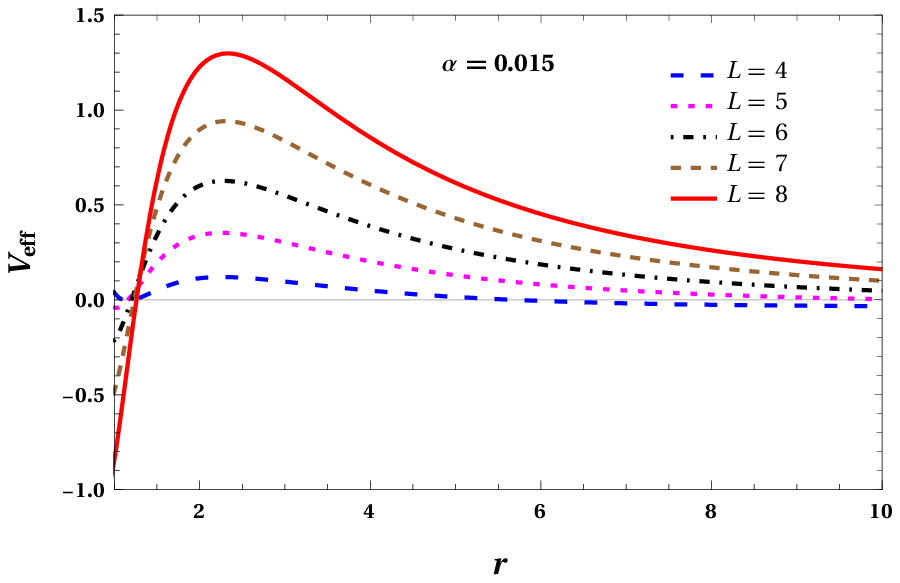}\label{Veff015b}}
\caption{The behaviour of effective potential $V_{eff}$ for 4D Gauss Bonnet black hole. In the left $L$ is within the range and on the right $L$ is outside the range. In each plot, the corresponding $a_E$ is taken.}
\label{Veffective}
\end{figure}

\begingroup
\setlength{\tabcolsep}{10pt} 
\renewcommand{\arraystretch}{1.2} 
\begin{center}
\begin{table}
\begin{tabular}{ |c|c|c|c|c|c|c| } 
\hline
$\alpha$ & $a$ & $r_H^-$&$r_H^+$ &$L_4$(min)&$L_3$(max)\\
\hline
\hline
0.005& 0.6 & 0.809234 &1.55524 &-6.25914&3.49874\\ 
0.0075& 0.5 & 0.813802 &1.53765 &-6.045&3.67612\\ 
0.01& 0.4 & 0.78801 &1.51457 &-5.81976&3.87095\\ 
0.0125& 0.3 & 0.748193 &1.47841 &-5.58007&4.07438 \\ 
0.015& 0.2 & 0.712647 &1.41926 &-5.32051&4.28064\\ 
\hline
\end{tabular}
\caption{\label{lrangenonext} The range of angular momentum of the infalling particle for the non-extremal rotating four dimensional Gauss Bonnet black hole.}
\end{table}
\end{center}
\endgroup
Since the geodesics are time like we must impose, $dt/d\tau >0$. Then, from the equation \ref{teqn} ,
\begin{equation}
\frac{1}{r^2}\left[ a(L-aE)+\frac{r^2+a^2}{\Delta}\left[ E(r^2+a^2)-aL\right] \right] \geq 0.
\end{equation}
This, under the limit $r-\rightarrow r_H^E$, reduces to
\begin{equation}
E-\Omega _H L\geq 0,
\end{equation}
with the angular velocity of the black hole on the horizon $\Omega _H$ as
\begin{equation}
\Omega _H=\frac{a}{r_H^E+a^2}.
\end{equation}
From this we have the critical angular momentum of the particle which is $L_c = E/{\Omega _H}$. The incoming particle with
critical angular momentum $L_c$ can approach the horizon of the black hole. If the angular momentum $L$ of the particle is above the critical value $L_c$ then it will never fall into the black hole. Whereas, if the angular momentum is below the critical value, the particle is always guaranteed to fall into the black hole.  This is clearly depicted in the behaviour of the effective potential (figure  \ref{Veffective}). When the angular momentum within the critical values the effective potential is negative, which corresponds to the bounded motion. On the other hand, for the angular momentum beyond the critical values, the effective potential is positive, and the motion is unbounded.

\section{Centre-of-mass energy in the rotating 4D Gauss Bonnet black hole}
\label{secthree}
\begin{figure}[H]
\centering
\subfigure[ref2][]{\includegraphics[scale=0.8]{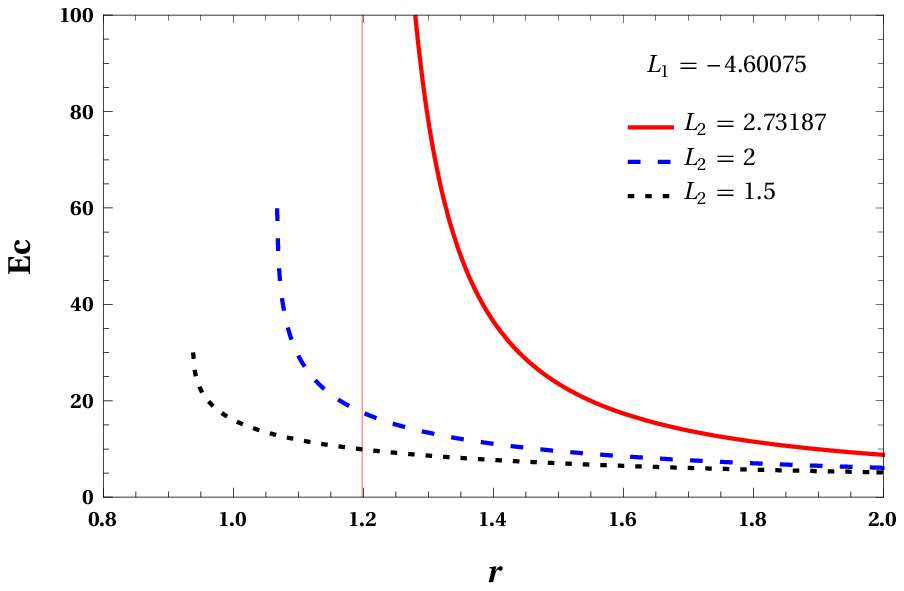}\label{Ecomext005}}
\qquad
\subfigure[ref1][]{\includegraphics[scale=0.8]{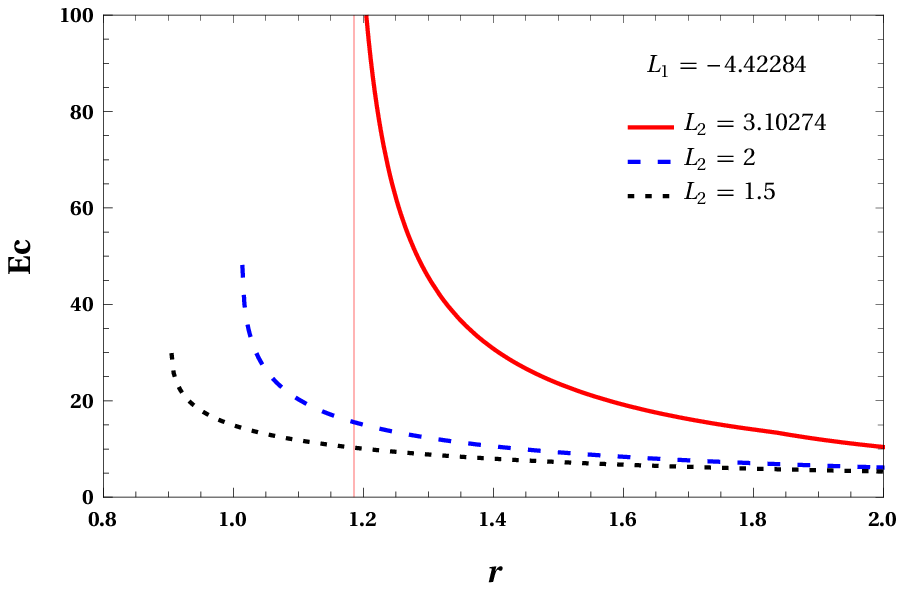}\label{Ecomext01}}
\caption{The behaviour of centre-of-mass energy for extremal black hole for different $\alpha$ values, $\alpha =0.005$ (left) and $\alpha =0.01$ (right). In both cases the corresponding $a_E$ and $r_H^E$ values are taken. The vertical line represents the event horizon.}
\label{Ecomext}
\end{figure}

\begin{figure}[H]
\centering
\subfigure[ref2][]{\includegraphics[scale=0.8]{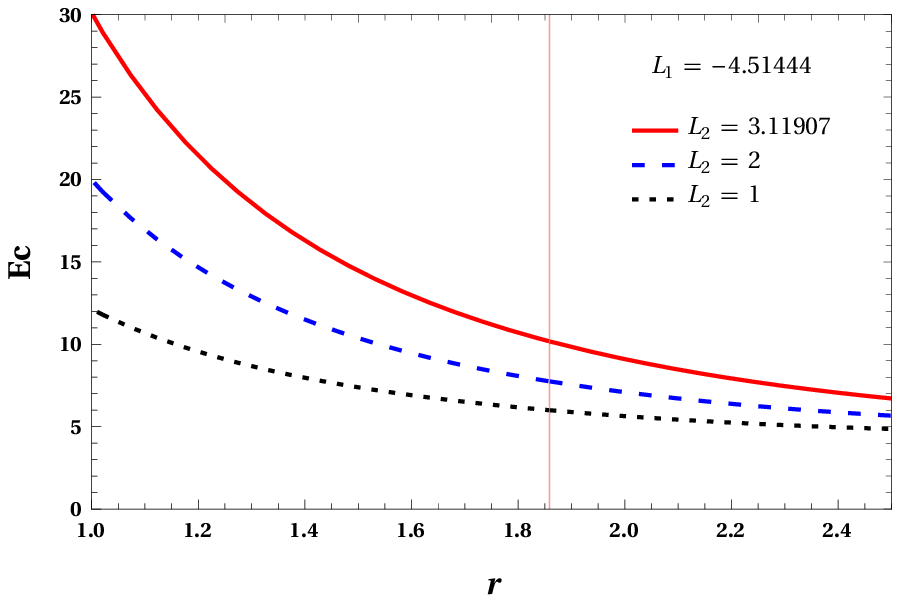}\label{Ecomnonext005}}
\qquad
\subfigure[ref1][]{\includegraphics[scale=0.8]{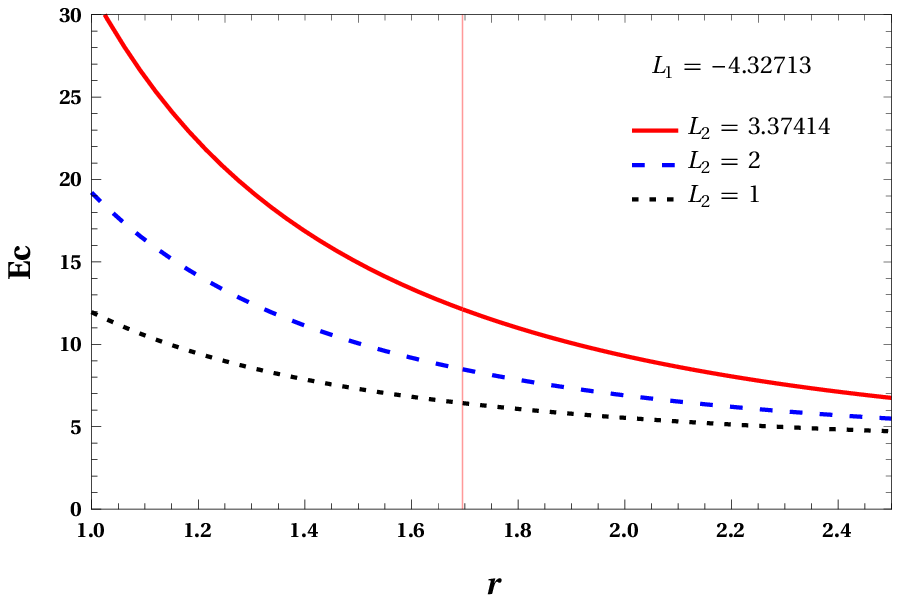}\label{Ecomnonext01}}
\caption{The behaviour of centre of mass energy for non-extremal black hole for different $\alpha$ values. We have taken $\alpha =0.005$ with $a=0.1$ (left) and $\alpha =0.01$ (right) with $a=0.1$. In both cases the corresponding $r_H^E$ values are taken. The vertical line represents the event horizon.}
\label{Ecomnonext}
\end{figure}

In this section we study the collision of two particles in the vicinity of the horizon of the rotating 4D Gauss Bonnet black hole in the equatorial plane. Initially the two particles are at rest at infinity ($E_1/m_1=1$ and $E_2/m_2=1$) and then they approach the black hole and collide at a distance $r$. We consider two particles of same mass $m_1=m_2=\mu$ and with different angular momenta $L_1$ and $L_2$. These angular momenta must lie in the range of values we calculated in the previous section, for the collision to takes place at the horizon of the black hole.  The collision energy of the particles in the centre-of-mass frame in the background of a rotating 4D Gauss Bonnet black hole is \cite{Banados:2009pr}
\begin{equation}
E_{CM}=\sqrt{2}\mu \sqrt{1-g_{\mu \nu} u^\mu _1 u^\nu _2}
\end{equation}
where $u_i^\mu=dx_i^\mu/d\tau$ ($i=1,2$) are the four velocities of the two particles. A simple calculation yields the following expression for the centre of mass energy,

\begin{eqnarray}
E^2_{CM}=\frac{2\mu ^2}{\Delta  r^2} \left[ (r^2+a^2)^2-a(L_1+L_2)(r^2+a^2-\Delta)+L_1L_2(a^2-\Delta)+\Delta(r^2-a^2)-X_1X_2\right]\nonumber \\
\end{eqnarray}

\begin{equation}
X_i=\sqrt{\left(a L_i-r^2-a^2\right)^2-\Delta  \left((L_i-a)^2+\mu ^2 r^2\right)} \quad (i=1,2)
\end{equation}
The above expression, which is invariant under the interchange $L_1 \leftrightarrow L_2$, confirms that the coupling parameter $\alpha$ has a significant influence on the $E_{CM}$, as $\Delta $ has the $\alpha$ dependence. We study the properties of centre-of-mass energy as the radius r approaches to the horizon $r_+$ for both the extremal and non-extremal black holes (figure \ref{Ecomext} and \ref{Ecomnonext}). For the collision to takes place we must ensure that the angular momenta of the colliding particles are within the allowed range. For the extremal black hole, the critical angular momentum is within the range whereas for the non-extremal case it lies outside the range. Only for the extremal black hole a particle can approach the horizon with the critical angular momentum.

 For the extremal black hole, the centre-of-mass energy $E_{CM}$ diverges when the angular momentum of one of the incoming particle has the critical value ($L_1=L_c$ or $L_2=L_c$) (the results are shown in figure \ref{Ecomext}). From this, we can say that the extremal rotating 4D Gauss-Bonnet black hole can act as a particle accelerator. The arbitrarily high centre-of-mass energy of the colliding particles at the horizon may provide an effective way to investigate the Planck-scale physics in the background of this extremal black hole. For the non-extremal black hole, the centre-of-mass energy has a finite upper limit (figure \ref{Ecomnonext}).



\section{Discussions}
\label{secfour}
Recently proposed four-dimensional black hole solutions provides a novel approach to understand Gauss-Bonnet gravity in lower dimensions. In this paper, we demonstrate that the rotating four-dimensional Gauss-Bonnet black hole can act as a particle accelerator. First, we studied the influence of the Gauss-Bonnet coupling parameter, the horizon structure and the static limit surface of the black hole. The coupling parameter imparts a different ergo sphere structure to the black hole compared to that of Kerr black hole. The existence of the black hole solution requires certain constraints on the value of coupling parameter $\alpha$ and the spinning parameter $a$. We studied the extremal and non-extremal cases of the black hole for different combinations of $\alpha$ and $a$.

Adopting the BSW mechanism, we have studied the properties of the centre-of-mass energy for two general particles colliding in the vicinity of the black hole. For that purpose, we have focused on the first order geodesic equations of a particle. Using the particle motion equations, the range of allowed angular momenta to reach the black horizon is obtained. Considering the incoming particles within this range, the particle collision is observed. Our results show that extremal black holes can serve as particle accelerators. The centre-of-mass energy has arbitrarily high value when one of the particle approaches with a critical angular momentum. However, for non-extremal black holes there is always an upper limit for the centre-of-mass energy. We have shown that the BSW mechanism depends on the value of Gauss-Bonnet coupling parameter $\alpha$ and the spinning parameter $a$ of the black hole.


\acknowledgments
Author N.K.A., A.R.C.L., and K.H. would like to thank U.G.C. Govt. of India for financial assistance under UGC-NET-SRF scheme.


  \bibliography{BibTex}

\end{document}